# Design and Implementation A different Architectures of mixcolumn in FPGA

Sliman Arrag[1], Abdellatif Hamdoun [2], Abderrahim Tragha [3] and Salah eddine Khamlich [4]

[1] Department of Electronics and treatment of information
UNIVERSITE HASSAN II MOHAMMEDIA, Casablanca, Morocco
Email arragsliman@yahoo.fr

[2] Department of Electronics and treatment of information
UNIVERSITE HASSAN II MOHAMMEDIA, Casablanca, Morocco
Email alhamdoun@yahoo.fr

[3] Department of computing and Mathematics
UNIVERSITE HASSAN II MOHAMMEDIA, Casablanca, Morocco
Email a.tragha@univh2m.ac.ma

[4] Department of Electronics and treatment of information
UNIVERSITE HASSAN II MOHAMMEDIA, Casablanca, Morocco
Email khamlich.salah@gmail.com

*ABSTRACT*

*This paper details Implementation of the Encryption algorithm AES under VHDL language In FPGA by using different architecture of mixcolumn. We then review this research investigates the AES algorithm in FPGA and the Very High Speed Integrated Circuit Hardware Description language (VHDL). Altera Quartus II software is used for simulation and optimization of the synthesizable VHDL code. The set of transformations of both Encryptions and decryption are simulated using an iterative design approach in order to optimize the hardware consumption. Altera Cyclone III Family devices are utilized for hardware evaluation.*

## 1. INTRODUCTION

The cryptographic algorithms became the main proceeding for protection of very important data, the security objective called confidentiality [1-3] being the one taken into account by their hardware implementation and by their integration into the present-day communication systems.

A number of the encryption/decryption algorithm the cryptographie has been developed [2-4]. Keeping pace with maturity of the security technology the hackers, the electronic eavesdroppers, electronic frauds and the virus have been coming into the field with new improved techniques for to attack the security mechanism [17], [19]. So to protect any attack to the valuable information source and their transmission, the algorithm Advanced Encryption Standard (AES or Rijndael), a Federal Information Processing Standard is approved by National Institute of Standards and Technology (NIST) [4], [7], [8], [11].But AES has 10 round of complex algebraic and matrix operation which involve high processing power and introduce delay in encryption and decryption process. For this reason at the beginning of this work the speed is treated as a major issue and concentration is provided on hardware based implementation. Field Programmable Gate Array based implementation is chosen in this operates as FPGA offers lower cost, flexibility and reasonable performance than ASIC (Application Specific Integrated Circuit) implementation. Beforehand researcher proposed application of AES processor on FPGA hardware place a few security features since earlier version of the FPGA available in the market was low capacity.

Newly the development of a AES CPU using VHDL and its implementation on FPGA Xillinx without sacrificing any security feature of the algorithm is reported [22]. It offers many FPGA high capacities in various families. Literatures [10], [12], [13], [18], [20-21] describe design and implementation of AES processor in the FPGA platform.

This paper presents a hardware implementation for the AES (Advanced Encryption Standard) symmetric cryptographic algorithm, under VHDL programming language by using different architecture of Mixcolumn and a hardware simulation of the resulted ciphering & deciphering module.

## 2. THE BASIC CRYPTOGRAPHIC ARCHITECTURE

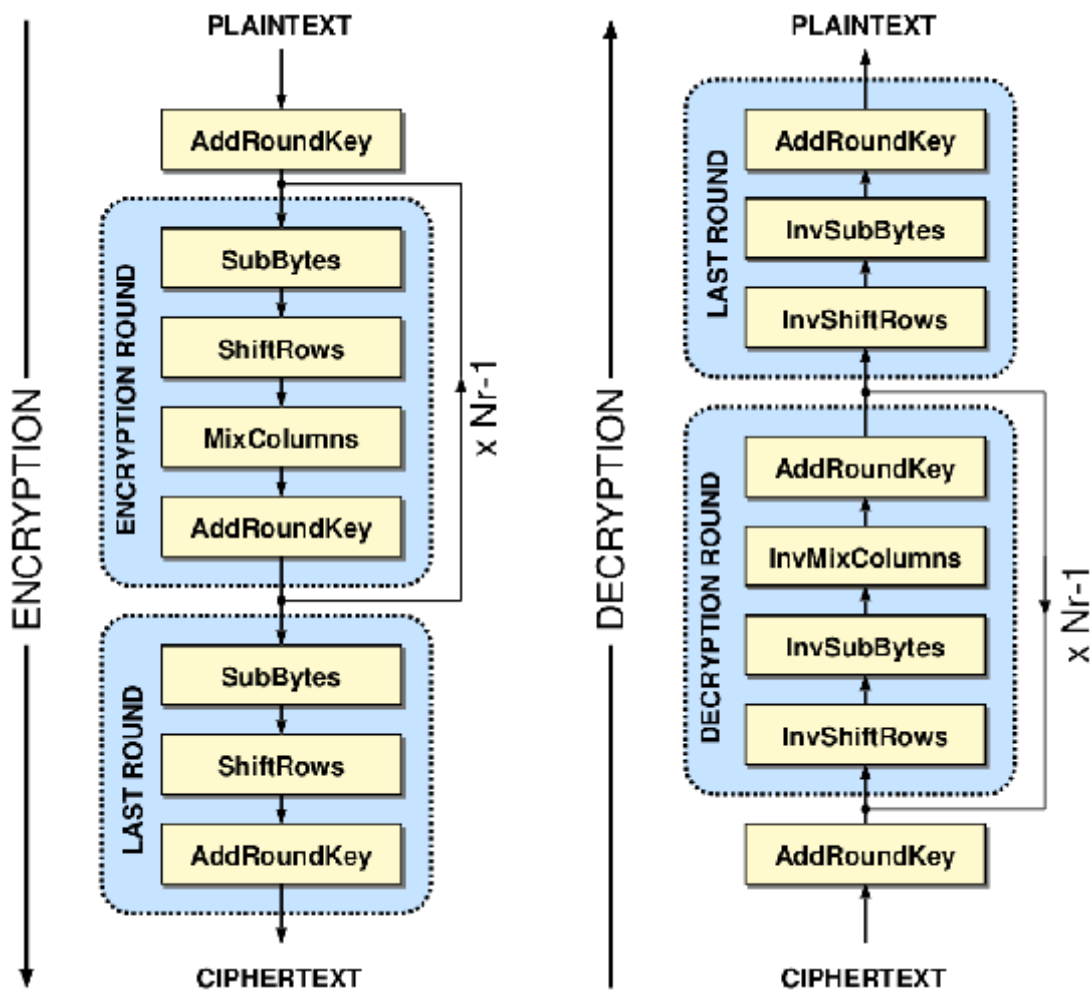

Figure 1. The basic AES-128 cryptographic architecture

AES algorithm is a FIPS standard and is a symmetric key [5], [9], in which the sender and recipient use only key for encryption and decryption. The data block length is fixed to be 128 bits (Nb = 4 words), while the length of the cipher key can be 128, 192 or 256 bits, and be represented by Nk = 4, 6, or 8 words respectively. Moreover, the AES algorithm is an iterative algorithm. The iterations are called rounds, and the total number of rounds, Nr is 10, 12, or 14, when the key length is 128, 192, or 256 bits, respectively. The 128 bit plaintext block is divided into 16 bytes. These bytes are mapped to a 4 x 4 array called the State, and all the internal operations of the AES algorithm are performed on the State. Each byte in the State is denoted by Si;j , (0 < i, j < 5) and is considered as an element of Galois Fields, GF(28). The irreducible polynomial used in the AES algorithm to construct, GF(28) field is

$$m(x) = x^8 + x^4 + x^3 + x + 1. \qquad (1)$$

In (Figure 1) AES encryption processes are presented. In the encryption of the AES algorithm, each round except for the final round consists of four transformations: the Sub__Bytes(), the Shift__Rows(), the Mix__Columns(), and the Add__RoundKey(), while the final round does not have the MixColumns() transformation.

The algorithm AES It can be cut in three blocks:

**Initial Round:** It is the first and simplest of the stages. it only counts one operation: Add Round Key.

Remark: The inverse of this operation bloc it is herself.

**N Rounds:** N being the number of iterations. This number varies according to the size of the key used. (128 bits need N=9, 192 bits need N=11 need 256 bits. N =13. This second stage is constituted of N iterations including each the four following operations: Sub Bytes, Shift Rows, Mix Columns, Add Round Key.

**Final Round:** This stage is nearly identical to one of the N iterations of the second stage. The only difference is that it doesn't include the operation Mix Columns

## 2.1. ADDROUN-key

In this transformation, a round key is added to the State by a simple bit wise XOR operation (that is a sum in Galois Fields). Each tower key consists of four words from the key schedule procedure

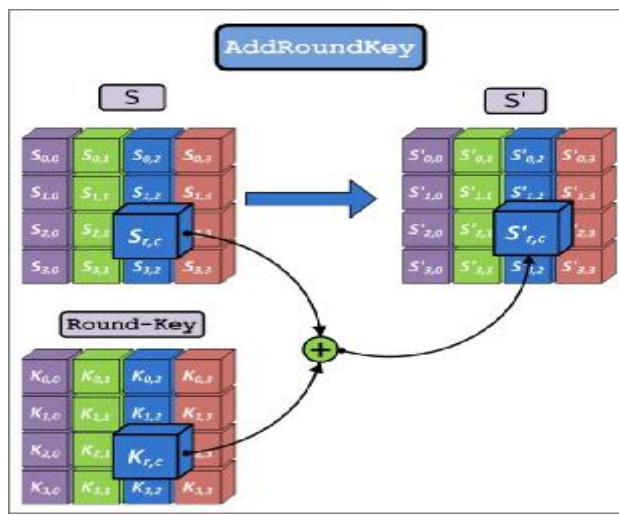

Figure 2.addround-key bloc

## 2.2. SubBytes

This transformation is a non-linear byte substitution that operates independently on each byte of the State using a substitution table (Sbox) [3]. We construct this S-box, which is reversible, by composing two transformations [9]:

- 1. Taking the multiplicative inverse in the finite field GF(28) with

    $$m(x) = x^8 + x^4 + x^3 + x + 1 \qquad (1)$$

- As irreducible polynomial; the element {00} is mapped onto itself.

2. Applying an affine (over GF (2) [1], [8]) transformation defined by:

$$b_i' = b_i \oplus b_{(i+4)\bmod 8} \oplus b_{(i+5)\bmod 8} \oplus b_{(i+6)\bmod 8} \oplus b_{(i+7)\bmod 8} \oplus c_i \quad (2)$$

- For $0 \leq i < 8$, where $b_i$ is the ith bit of the byte and $c_i$ is the ith bit of a constant byte c with the value {63}.

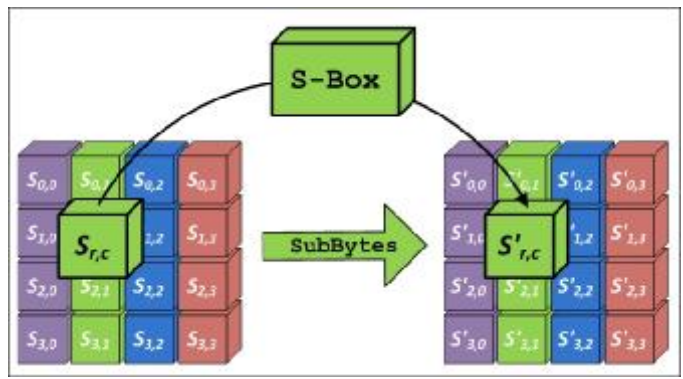

Figure 3. Sub-Byte bloc

- Remark: The function inverse named Inv_SubBytes consists in applying the same function but this time while using Inv_SBox that is the inverse table of SBox.

### 2.3. Shif-Rows

In this transformation, the bytes in the last three rows of the State are cyclically shifted over different numbers of bytes (offsets). The first row, row 0, is not shifted. Row 1 of the State is left shifted by 1 byte position; row 2 is left shifted by 2 byte positions; row 3 is left shifted by 3 byte positions.

- Remark: The inverse function of this Inv_ShiftRoxs operation consists in replacing the shift on the left on the right by a shift.

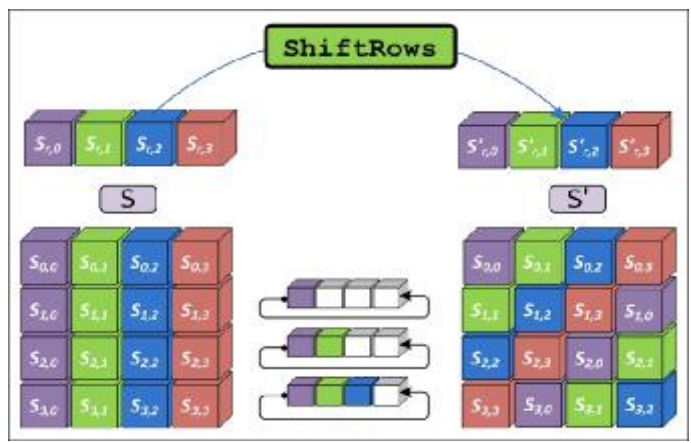

Figure 4. Shift-row bloc

## 2.4. MixColumns

This transformation [9] operates on the State column-by-column, treating each column as a four-term polynomial over GF(28). *These polynomials are* multiplied modulo (x4 + 1) with a fixed polynomial a(x), specified in the standard.

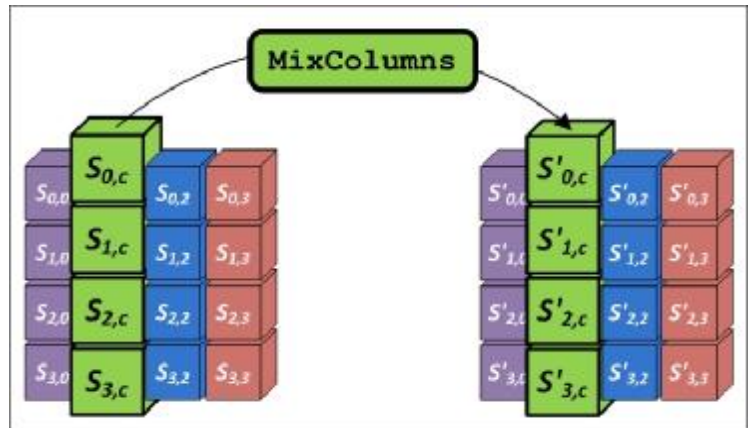

Figure 5. Mix-column bloc

| 02 | 03 | 01 | 01 | D4 | E0 | B8 | 1E |
|----|----|----|----|----|----|----|----|
| 01 | 02 | 03 | 01 | BF | B4 | 41 | 27 |
| 01 | 01 | 02 | 03 | 5D | 52 | 11 | 98 |
| 03 | 01 | 01 | 02 | 30 | AE | F1 | E5 |

Figure 6. Exemple of Multiplication the Mixcolumn

*Result [1,1] = 02•D4 XOR 03•BF XOR 01•5D XOR 01•30*

- Remark: The inverse function of this operation, InvMixColumns, consists in multiplying the State_out matrix by the inverse of the constant matrix.

$$\begin{pmatrix} 0E & 0B & 0D & 09 \\ 09 & 0E & 0B & 0D \\ 0D & 09 & 0E & 0B \\ 0B & 0D & 09 & 0E \end{pmatrix}$$

Figure 7. Matrice of Multiplication inverse the Mixcolumn

**Previous work**

There exist many presentations of hardware implementations of AES algorithms in literature. In 2001, Elbirt et al., [6] compared five candidate algorithms for AES using Field-Programmable Gate Array (FPGA) implementations. Later FPGA implementations demonstrate better utilization of FPGA resources. Several architectures using dedicated on-chip memories implementing S-boxes and Mix-column were developed [23] [24] [25] [26] [27]. Recent research focused on fast pipelined, paralleled and optimal power calculation for the cryptography AES implementations in both FPGA [13] [28] [29] [30] [31] [32]. Unfortunately, most of those implementations are too costly for practical applications.

In this paper, we have developed compared and implemented our new different architectures of mix-column:

1-Methods permit to calculate the products in GF (28) (architecture (1).
2- Galois Multiplication lookup tables (architecture 2).
3- Properties of the binary calculation (architecture 3).
To achieve a high throughput with small area. The rest of the paper is organized as follows:
Section 3. Implementation of AES (Ciphering & Deciphering) in FPGA.
Section 4. simulation & interpretation.
Section 5 concludes the paper.

### *2.4.1. methods permit to calculate the products in GF (28).*

a) *Mathematical application (architecture 1)* [14]

$\{0D\}* M(x) = (m0 + m5 + m6) + (m1 + m5 + m7) x + (m0 + m2 + m6) x2 + (m0 + m1 + m3 + m5 + m6 + m7) x3 + (m1 + m2 + m4 + m5 + m7) x4 + (m2 + m3 + m5 + m6) x5 + (m3 + m4 + m6 + m7) x6 + (m4 + m5 + m7) x7.$ \hfill (3)

$\{0E\}* M(x) = (m5 + m6 + m7) + (m0 + m5) x + (m0 + m1 + m6) x2 + (m0 + m1 + m2 + m5 + m6) x3 + (m1 + m2 + m3 + m5) x4 + (m2 + m3 + m4 + m6) x5 + (m3 + m4 + m5 + m7) x6 + (m4 + m5 + m6) x7.$ \hfill (4)

$\{0B\}* M(x) = (m0 + m5 + m7) + (m0 + m1 + m5 + m6 + m7) x + (m1 + m2 + m6 + m7) x2 + (m0 + m2 + m3 + m5) x3 + (m1 + m3 + m4 + m5 + m6 + m7) x4 + (m2 + m4 + m5 + m6 + m7) x5 + (m3 + m5 + m6 + m7) x6 + (m4 + m6 + m7) x7.$ \hfill (5)

$\{09\}* M(x) = (m0 + m5) + (m1 + m5 + m6) x + (m2 + m6 + m7) x2 + (m0 + m3 + m5 + m7) x3 + (m1 + m4 + m5 + m6) x4 + (m2 + m5 + m6 + m7) x5 + (m3 + m6 + m7) x6 + (m4 + m7) x7.$ \hfill (6)

$\{03\}* M(x) = (m0 + m7) + (m0 + m1 + m7) x + (m1 + m2) x2 + (m2 + m3 + m7) x3 + (m3 + m4 + m7) x4 + (m4 + m5) x5 + (m5 + m6) x6 + (m6 + m7) x7.$ \hfill (7)

$\{02\}* M(x) = (m7) + (m0 + m7) x + (m1) x2 + (m2 + m7) x3 + (m3 + m7) x4 + (m4) x5 + (m5) x6 + (m6) x7.$ \hfill (8)

$\{01\}* M(x) = M(x)$ \hfill (9)

b) *Galois Multiplication lookup tables (architecture 2)*

Commonly, rather than implementing galois multiplication, Rijndael implementations simply use pre-calculated lookup tables to perform the byte multiplication by 02, 03, 09, 0B, 0D, and 0E.[15].
These lookup tables are as follows:

*Multiplication by 02*

```
0x00,0x02,0x04,0x06,0x08,0x0a,0x0c,0x0e,0x10,0x12,0x14,0x16,0x18,0x1a,0x1c,0x1e,
0x20,0x22,0x24,0x26,0x28,0x2a,0x2c,0x2e,0x30,0x32,0x34,0x36,0x38,0x3a,0x3c,0x3e,
0x40,0x42,0x44,0x46,0x48,0x4a,0x4c,0x4e,0x50,0x52,0x54,0x56,0x58,0x5a,0x5c,0x5e,
0x60,0x62,0x64,0x66,0x68,0x6a,0x6c,0x6e,0x70,0x72,0x74,0x76,0x78,0x7a,0x7c,0x7e,
0x80,0x82,0x84,0x86,0x88,0x8a,0x8c,0x8e,0x90,0x92,0x94,0x96,0x98,0x9a,0x9c,0x9e,
0xa0,0xa2,0xa4,0xa6,0xa8,0xaa,0xac,0xae,0xb0,0xb2,0xb4,0xb6,0xb8,0xba,0xbc,0xbe,
0xc0,0xc2,0xc4,0xc6,0xc8,0xca,0xcc,0xce,0xd0,0xd2,0xd4,0xd6,0xd8,0xda,0xdc,0xde,
0xe0,0xe2,0xe4,0xe6,0xe8,0xea,0xec,0xee,0xf0,0xf2,0xf4,0xf6,0xf8,0xfa,0xfc,0xfe,
0x1b,0x19,0x1f,0x1d,0x13,0x11,0x17,0x15,0x0b,0x09,0x0f,0x0d,0x03,0x01,0x07,0x05,
0x3b,0x39,0x3f,0x3d,0x33,0x31,0x37,0x35,0x2b,0x29,0x2f,0x2d,0x23,0x21,0x27,0x25,
0x5b,0x59,0x5f,0x5d,0x53,0x51,0x57,0x55,0x4b,0x49,0x4f,0x4d,0x43,0x41,0x47,0x45,
0x7b,0x79,0x7f,0x7d,0x73,0x71,0x77,0x75,0x6b,0x69,0x6f,0x6d,0x63,0x61,0x67,0x65,
0x9b,0x99,0x9f,0x9d,0x93,0x91,0x97,0x95,0x8b,0x89,0x8f,0x8d,0x83,0x81,0x87,0x85,
0xbb,0xb9,0xbf,0xbd,0xb3,0xb1,0xb7,0xb5,0xab,0xa9,0xaf,0xad,0xa3,0xa1,0xa7,0xa5,
0xdb,0xd9,0xdf,0xdd,0xd3,0xd1,0xd7,0xd5,0xcb,0xc9,0xcf,0xcd,0xc3,0xc1,0xc7,0xc5,
0xfb,0xf9,0xff,0xfd,0xf3,0xf1,0xf7,0xf5,0xeb,0xe9,0xef,0xed,0xe3,0xe1,0xe7,0xe5,
```

Figure 8. Lookup table of multiplication by 02

Note: Multiplication by 03, 09, 0B, OD and 0E [15]

*c) Properties of the binary calculation (architecture 3)*

According to the two architectures previous we were able to achieve other method based on the Properties of the binary calculation that have for goal the easiness the use of this operation to the material level you find the manner and the stages that we followed in order to calculate the multiplication mixcolumn below:

Multiplication by 01 (00000001 in binary): The number remains unaltered

Multiplication by 02 (00000010 in binary): The bits of the number are baffled toward the left:

N = 10101110 => 2N = 101011100

Since the operations make themselves in a number finished of the values (field of Galois GF (28) of 256 values), the MSB of 2N must be omitted:

2N = 101011100

2N = 01011100

If the MSB was (that we have just omitted) a '0', then 2N are the final result of the multiplication:

2N = 01011100

If the MSB was (that we have just omitted) a '1', what is the case in this example, then it is necessary to add (XOR) the binary number again 00011011 (1B) to 2N in order to compensate the loss of the MSB caused (provoked) by the shift:

2N XOR 00011011 = 01011100 XOR 00011011 = 01000111 2N = 01000111

For the Multiplications (03; 09; 0B; 0D; 0E):

we take The MSB like a mask and one calculates the operations (temp1, temp2 and temp3) of shift on the left to add by the number (1B) in order to compensate the loss of the MSB caused (provoked) by the shift.

and_mask := m(7) & m(7) & m(7) & m(7) & m(7) & m(7) & m(7) & m(7);

temp1:= m(6 downto 0) & '0' xor (("00011011") and and_mask);

and_mask := temp1(7) & temp1(7) & temp1(7) & temp1(7) & temp1(7) & temp1(7) & temp1(7) & temp1(7);

temp2:= temp1(6 downto 0) & '0' xor (("00011011") and and_mask);

and_mask := temp2(7) & temp2(7) & temp2(7) & temp2(7) & temp2(7) & temp2(7) & temp2(7) & temp2(7);

temp3:= temp2(6 downto 0) & '0' xor (("00011011") and and_mask);

Multiplication par 03 (00000011 en binaire) :temp1 xor m

Multiplication par 09 (00001001 en binaire) : temp3 xor m

Multiplication par 0B (00001011 en binaire) : temp1 xor temp3 xor m

Multiplication par 0D (00001101 en binaire) : temp2 xor temp3 xor m

Multiplication par 0E (00001110 en binaire) : temp1 xor temp2 xor temp3

Note => &: Operators of concatenation

## 3. Implementation of AES (Ciphering & Deciphering) in FPGA:

To concretize that our modifications give a better results in terms of area and speed than the previous work, we compare the encryption /decryption codes (original and modified) based on the three models of mix-column. The comparison considered two criteria: chip speed and area utilization. The design was implemented on an Cyclone III (EP3C80F780C6 model) device. Both designs were synthesized using Quartus II v9.1 tool. As shown in Table 1, an increase in speed about 20% was achieved in our design and a reduction in area about 12% was achieved in our design. On the other side, since we use Properties of the binary calculation to perform Mix-Column transformation, we the other encryption (model 1 and 2) design needs more memory (as shown in Table 1).

Implementation uses the VHDL programming language that nowadays is commonly a language used very established for FPGA [16]. The drawing & the software of the simulation is Quartus II.

The encryption block is represented in Figure 9, where the main signals used by the implementation are shown.

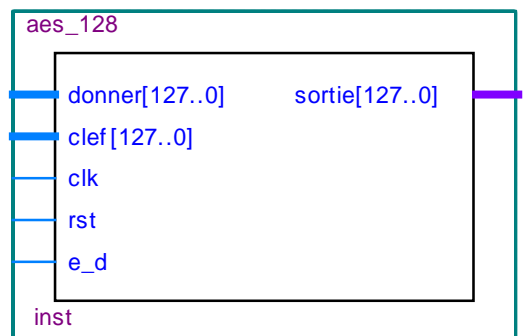

Figure 9. AES (ciphering & decoding) block

Table1. Comparative table between different blocks constitute AES algorithm

| Implementation | FPGA Device | | | | |
| --- | --- | --- | --- | --- | --- |
| | Total pins | logic cells | Peak virtual memory Megabyte | Total registers | Total memory bits |
| SBOX /invSBOX | 18 | 9 | 195 | 1 | 2048 |
| Bytsub/Invbytesub | 285 | 129 | 203 | 1 | 32768 |
| Shiftrow/Invshiftrow | 285 | 128 | 183 | 128 | 0 |
| Mixcolumn_arch1 | 285 | 224 | 183 | 128 | 0 |
| Mixcolumn_arch2 | 285 | 200 | 183 | 128 | 0 |
| Mixcolumn_arch3 | 285 | 200 | 195 | 128 | 0 |
| InvMixcolumn_arch1 | 285 | 372 | 185 | 128 | 0 |
| InvMixcolumn_arch2 | 285 | 959 | 206 | 128 | 0 |
| InvMixcolumn_arch3 | 285 | 360 | 183 | 128 | 0 |
| Generation_de_clef | 261 | 992 | 191 | 0 | 0 |
| Aes_128_arch1 | 387 | 75840 | 370 | 128 | 0 |
| Aes-128_arch2 | 387 | 80829 | 582 | 128 | 0 |
| Aes_128_arch3 | 387 | 75147 | 368 | 128 | 0 |

we take in consideration the different architecture (model 1,2 and 3) of Block Mixcolumn that we already treated, under VHDL language in the Circuit Cyclone III (EP3C80F780C6 model) which is a low resource: 81264 logical elements; 430 pine to enter / exit; Embedded Multiplier 9-bit elements 488; capacity of memory 2810880bits; 4PLL.The main signals are: the clock of the system(CLK), the system reset (RST), signal of the load which load the key and given them and signal that permits to encode and to decipher given them. A summary of the occupied resources is presented in the comparative table. (Table 1)

• According to the comparative table we can notice that with first architecture of AES-128 (based on the model 1 of mixcolumn) of the setting in .implementation occupies more that (75840 slices) of the device, when the second (based on the model 3 of mixcolumn) need of the setting in .implementation roughly (75147tranches) of capacity total of the device, on the other hand the last architecture of AES-128 has (based on the model 2 of mixcolumn) need (80829) of the device.

The conclusion is that the last (model 3) bet in implementation is more efficient than architecture of the first and the third, about the number of occupation of resources of the device.

## 4. SIMULATION & INTERPRETATION

Each round has 4 operations and it is iterative in nature. So the output of first round is fed to the second round as input data and performs the same operations with another set of keys. This process continued until the last round reach. In the last round, there is no mix-column operation. The State array obtained after the last round is the required cipher text for transmission

Figure 10 and 11: shows the simulation results of the encrypt and decrypt data, we give 1bit e_d (to control the encryption and decryption ,e_d=1 Ciphering else Deciphering ), 128 bit plaintext (data that we need to encrypt) and 128 bit key (the key used also to generate another keys), of course clock (clk) was used to synchronize the various blocks and rst representing the reset the given if equals 1, finally, we will have in 128 bit output (cyphertext).we use for our simulation QuartusII simulator.

The diagrams retailed of the simulation the processes for the setting in implementation AES are presented below, in Figure (10 and 11).The length total of the process of the ciphering is (124s) and some decoding is (277s).

**Ciphering** :( figure10)

**e_d**:1

**Plaintext**: hamdoun_&_tragha

**Key:** arragsliman_miti

**cyphertext**:8[139][195]S[189] :[190]P[206][221][153][132][205]bI*

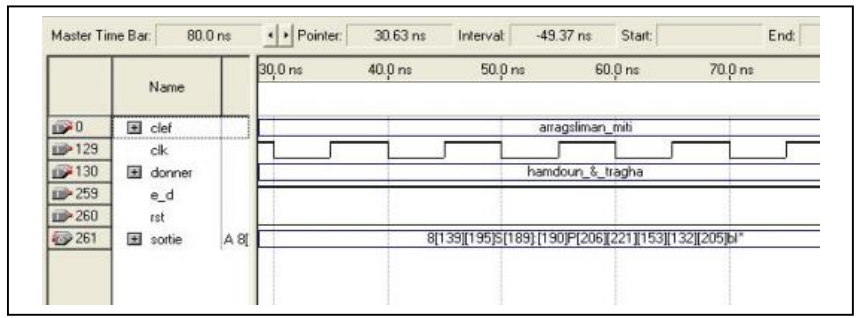

Figure 10.  Simulation of the ciphering of AES-128

**Deciphering** :( figure 11)

**e_d**:0

**plaintext** :8[139][195]S[189] :[190]P[206][221][153][132][205]bI*

**Key**: arraglsiman_miti

**Cyphertext**: hamdoun_&_tragha

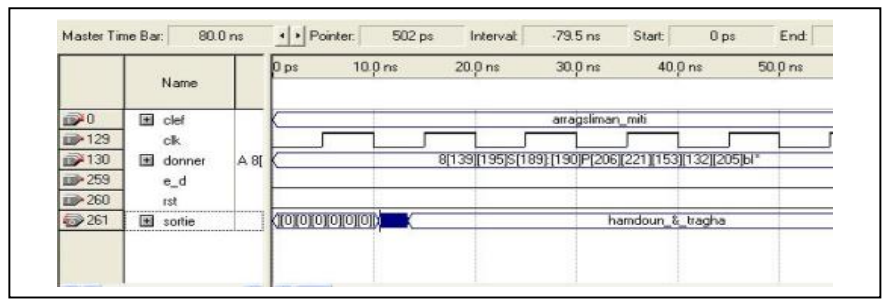

Figure 11.  Simulation of the decoding of AES-128

During our implementation, we met several difficulties among which we mention:

- Differentiation in time of execution between the ciphering and the decoding because of the structure of some functions and we could remedy this problem by optimization of the code

- The problems of battery overflow caused by the operating system, and we think that it is preferable to work with machines of big performances (RAM, speed of clock, cache memory.).

- Calculate it binary in field of Galois notably the multiplication and the inverses of the matrixes.

- The description sometimes dark of the algorithm of RIJNDAEL and especially in the phase of decoding.

This system of cryptage is used in the domains industrial, commercial and financial and on an increasing number of the PCs. its domination on the market increases daily.

## 5. CONCLUSIONS

In this paper, we have presented a novel FPGA implementation of the encryption algorithm AES under VHDL utilizing high performance Mix-column/inv-Mix-column which uses Properties of the binary calculation.

The result shows the FPGA implementations allow us to increase flexibility, lower costs, and reduce time to release enhanced cryptographic equipment, providing a satisfactory level of security for communication applications, or other electronic data transfer processes where security is needed.

The modification (AES by using architecture 3 of Mix-column) of gives an 12% reduction in area and 20% increase in speed compared with the original design (AES by using architectures 1 and 2 of Mix-column) . Our design gives the highest throughput and area utilization over all the Iterative Looping based FPGA implementations. The decryption algorithm is implemented and gives better results than the design in previous work.

This compact design can help in implementing AES for smart cards, RFID Tags, and wireless sensors. This design prevents timing attack on mix-columns as the resultant columns take the same duration not depending on multiplicand.

Our optimized and Synthesizable VHDL code is developed for the implementation of both encryption and decryption process. Each program is tested with some of the sample vectors provided by NIST and output results are perfect with minimal delay. Therefore, AES can indeed be implemented with reasonable efficiency on an FPGA, with the encryption and decryption taking an average of 124 and 277(s) respectively (for every 128 bits). The time varies from chip to chip and the calculated delay time can only be regarded as approximate.